\documentclass[prl,twocolumn,showpacs,superscriptaddress]{revtex4}
\usepackage{graphicx}
\usepackage{dcolumn}
\usepackage{amsmath}

\begin{document}
\title{The Orbital Order Parameter in La$_{0.95}$Sr$_{0.05}$MnO$_{3}$ probed by Electron Spin Resonance}
\author{J.~Deisenhofer}
\affiliation{Experimentalphysik V, Center for Electronic
Correlations and Magnetism, Institute for Physics, Augsburg
University, D-86135 Augsburg, Germany}
\author{B.I.~Kochelaev}
\affiliation{Kazan State University, 420008 Kazan, Russia}
\author{E.~Shilova}
\affiliation{Kazan State University, 420008 Kazan, Russia}
\author{A.M.~Balbashov}
\affiliation{Moscow Power Engeneering Institute, 105835 Moscow,
Russia}
\author{A.~Loidl}
\affiliation{Experimentalphysik V, Center for Electronic
Correlations and Magnetism, Institute for Physics, Augsburg
University, D-86135 Augsburg, Germany}
\author{H.-A.~Krug von Nidda}
\affiliation{Experimentalphysik V, Center for Electronic
Correlations and Magnetism, Institute for Physics, Augsburg
University, D-86135 Augsburg, Germany}

\date{\today}

\begin{abstract}
The temperature dependence of the electron-spin resonance
linewidth in La$_{0.95}$Sr$_{0.05}$MnO$_{3}$ has been determined
and analyzed in the paramagnetic regime across the orbital
ordering transition. From the temperature dependence and the
anisotropy of linewidth and $g$-value the orbital order can be
unambiguously determined via the mixing angle of the wave
functions of the $e_{\rm g}$-doublet. The linewidth shows a
similar evolution with temperature as resonant x-ray scattering
results.
\end{abstract}


\maketitle

In transition-metal oxides the orbital degrees of freedom play an
important role for the electric and magnetic properties. Their
coupling to spin, charge and lattice is responsible for the
occurrence of a variety of complex electronic ground states.
Orbital order (OO) can be derived via the Jahn-Teller (JT) effect
or via superexchange (SE) between degenerate orbitals under the
control of strong Hund's-rule coupling \cite{Jahn37}. Strong
correlations exist between spin and orbital order and between OO
and lattice distortions, but of course a one-to-one correspondence
cannot be expected. While spin and lattice order can easily be
detected experimentally, this is not true for OO and so far the OO
parameter remains hidden. In recent years resonant x-ray
scattering (RXS) has been used to derive information on the OO
parameter \cite{Murakami98b}, but there is an ongoing dispute,
whether RXS probes the JT distortion or the orbital charge
distribution \cite{Dispute,Ishihara02}. Indirectly, OO can also be
derived from diffraction experiments via lattice distortions and
bond lengths \cite{Rodriguez-Carvajal98}. In this Letter we
demonstrate that electron-spin resonance (ESR) can be used to
detect OO and to monitor the evolution of the OO parameter.
Probing the spin of the partially filled $d$-shell of Mn$^{3+}$
ions by ESR, the anisotropy and $T$-dependence of $g$-value and
linewidth $\Delta H$ provide clear information on OO via
spin-orbit (SO) coupling.

The power of ESR to gain insight into OO will be demonstrated on
A-type antiferromagnetic (AFM) LaMnO$_3$ ($T_{\rm N}$=140~K), the
parent compound of the magneto-resistance manganites and a
paradigm for a cooperative JT effect that suggests a
$d_{3x^{2}-r^{2}}$/$d_{3y^{2}-r^{2}}$-type OO below $T_{\rm
N}$=750~K \cite{Goodenough55}. However, it has been shown that SE
interactions play an important role, too \cite{Okamoto}. Several
recent studies exhibit clear anomalies of the ESR parameters at
the JT transition in both doped and pure LaMnO$_3$
\cite{Huber99,Tovar99,deisenhofer02,Ivanshin00}. The orbitally
ordered $O^{\prime}$-phase is characterized by an anisotropy of
$\Delta H$ \cite{Ivanshin00,Alejandro01}, which for
polycrystalline samples reduces to a broad maximum in $\Delta
H$($T$) \cite{Tovar99,deisenhofer02}. Previously, the angular
dependencies of $\Delta H$ and the resonance field $H_{\rm res}$
had been analyzed for 200~K and 300~K in high-temperature
approximation, allowing to estimate the Dzyaloshinsky-Moriya (DM)
interaction and the strength of the zero-field splitting (ZFS)
parameters \cite{Deisenhofer02}. At X-Band frequencies (9 GHz)
$\Delta H$ was of the same order of magnitude as $H_{\rm res}$ and
due to the overlap with the resonance at $-H_{\rm res}$  and their
mutual coupling via the nondiagonal elements of the dynamic
susceptibility \cite{Benner83} the values for $H_{\rm res}$
contained a rather large uncertainty. To avoid these problems we
performed new experiments at Q-band frequencies (34 GHz), which
allowed a better determination of $H_{\rm res}$.

The scope of the present paper is the comprehensive analysis of
the $T$-dependence and anisotropy of $\Delta H$ and  the $g$-value
of the ESR signal in La$_{0.95}$Sr$_{0.05}$MnO$_{3}$. We chose
this concentration for the present study as an untwinned single
crystal was available and $T_{\rm JT}$$\sim$605~K is accessible to
our experimental setup \cite{Paraskevopoulos00}.

New ESR measurements were performed with a Bruker ELEXSYS E500 CW
spectrometer at Q-Band frequencies ($\nu \approx$ 34 GHz, 4.2~K
$\leq T \leq$ 290~K), using a continuous gas-flow cryostat for He
(Oxford). The oriented sample was mounted in a quartz tube with
paraffin. A goniometer allowed the rotation of the sample around
an axis perpendicular to the static magnetic field $\vec{H}_{\rm
ext}$.

\begin{figure}
\includegraphics[width=65mm,angle=0]{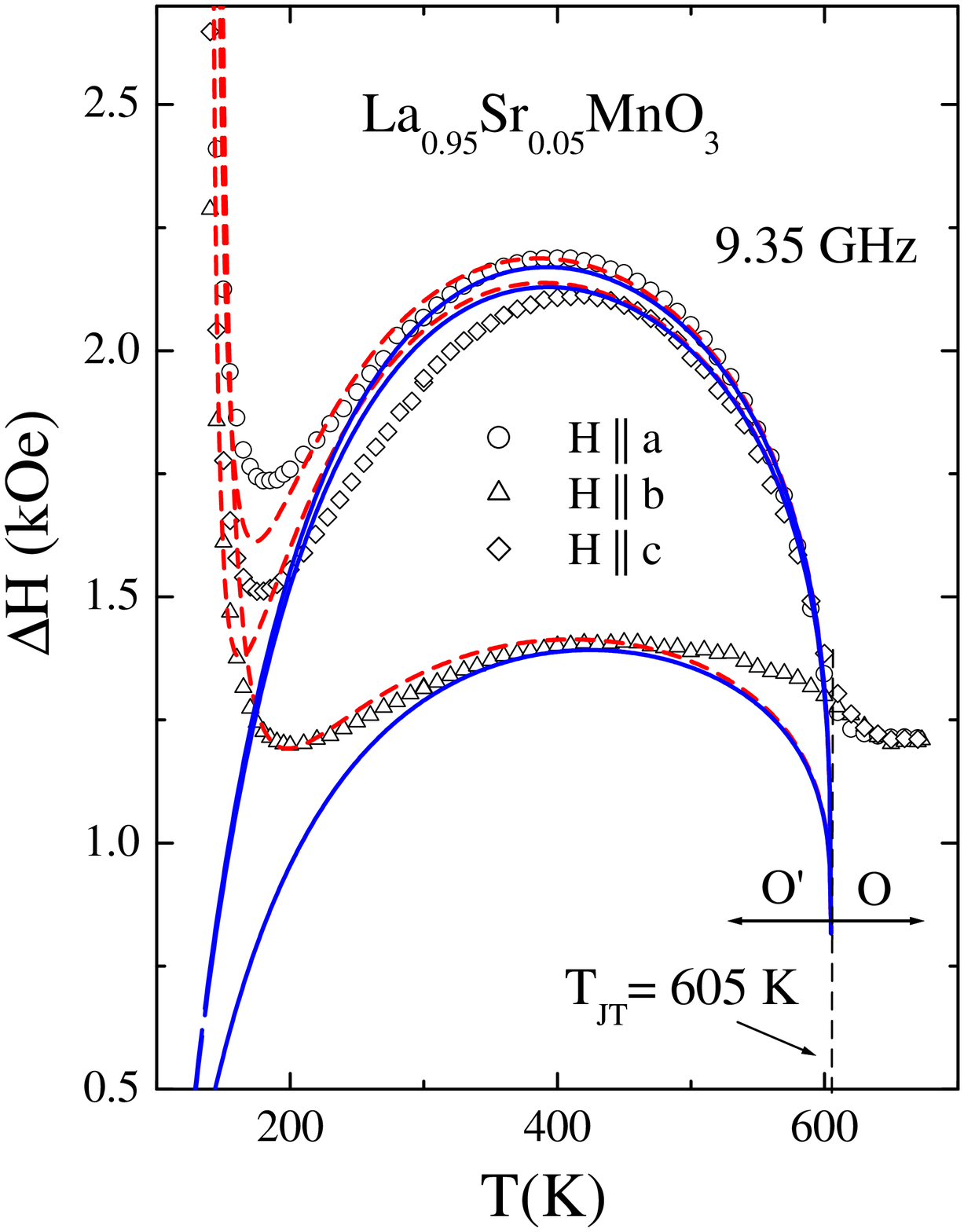}
\caption{$\Delta H$($T$) in La$_{0.95}$Sr$_{0.05}$MnO$_{3}$ at
X-Band frequency for $\vec{H}_{\rm ext}$ applied parallel to the
three main axes of the orthorhombic structure. Solid and dashed
lines represent fits using eq.~(\ref{dhfit}) as described in the
text.} \label{dhxband}
\end{figure}

\begin{figure}[t]
\includegraphics[width=60mm,angle=0]{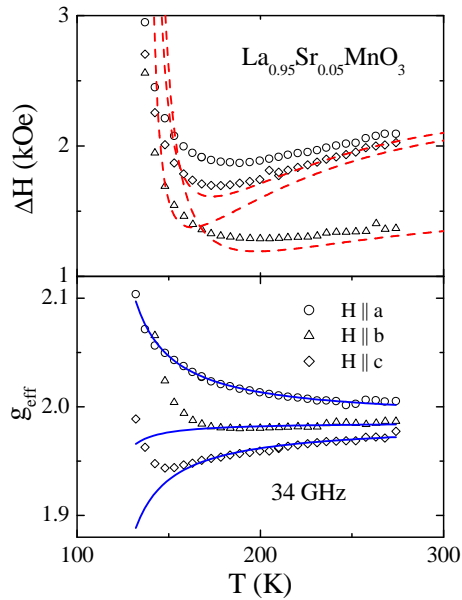}
\caption{$\Delta H (T)$ and effective g-value $g_{\rm eff}(T)$ at
Q-band frequency for $\vec{H}_{\rm ext}$ applied parallel to the
three crystallographic axes of the orthorhombic structure. The
solid lines represent fits using eq.~(\ref{gvalue}) for $g_{\rm
eff}$, the dashed lines for $\Delta H$ are the same as in
Fig.~\ref{dhxband}.} \label{qband}
\end{figure}

Figure \ref{dhxband} shows $\Delta H$ for X-band frequency and
Fig.~\ref{qband} $\Delta H$ and the effective $g$-value $g_{\rm
eff}$=$h\nu/(\mu_{\rm B}H_{\rm res})$ determined from $H_{\rm
res}$ for Q-band frequency. The observed linewidths at both
frequencies nicely coincide. Only near the minimum below 200~K the
absolute values are slightly enhanced at 34~GHz as compared to
9~GHz. Whereas the $g$-values obtained at X-band frequency bear a
rather large uncertainty \cite{Ivanshin00,Deisenhofer02}, at
Q-band frequency the $g$-values show a regular $T$-dependence,
approaching a constant high-temperature value and increasing for
$T \rightarrow T_{\rm N}$.
First, we determined the ZFS parameters $D$ and $E$ from the $T$
dependence of $H_{\rm res}$ at Q-band frequency. Using the general
formula for the resonance shift due to crystal field (CF) effects
\cite{Moreno01,Deisenhofer02} and accounting only for the rotation
(angle $\gamma$) of the MnO$_6$ octahedra in the $ac$ plane (axis
notation like in Ref.~\onlinecite{Huang97}), we obtained the
following expressions for the effective $g$-values for
$\vec{H}_{\rm ext}$ applied along one of the crystallographic axes
\begin{eqnarray}\label{gvalue}
\frac{g_{a,c}^{\rm eff}(T)}{g_{a,c}} &\approx& 1+\frac{D}{T-T_{\rm
CW}}\left[(3\zeta-1)\pm
3(1+\zeta)\sin(2\gamma)\right] \nonumber\\
\frac{g_{b}^{\rm eff}(T)}{g_b} &\approx& 1-\frac{2D}{T-T_{\rm
CW}}(3\zeta-1),
\end{eqnarray}
with the Curie-Weiss (CW) temperature $T_{\rm CW}$ and
$\zeta=E/D$. All terms of second and higher order in $D/(T-T_{\rm
CW})$ were neglected. The $T$-dependence of the effective
$g$-values is, hence, given by the CW law of the magnetic
susceptibility. Excluding the critical regime on approaching
magnetic order below 170~K, the data are well described by this
approach (solid lines in Fig.~\ref{qband}), where $T_{\rm CW}$ was
kept fixed at 111~K \cite{Ivanshin00}. The rotation angle was
chosen as $\gamma$=13$^{\circ}$ as observed in pure LaMnO$_3$
\cite{Huang97,Rodriguez-Carvajal98}. Then all data can be
consistently described by $D=0.60(2)$~K, the $E/D$-ratio
$\zeta$=0.37(1), and the $g$-values $g_a$=1.988(1),
$g_b$=1.986(1), and $g_c$=1.984(1). From these crystallographic
$g$-values, the local $g$ values of Mn$^{3+}$ can be calculated as
$g_z$=1.977, $g_y$(=$g_b$)=1.986, and $g_x$=1.995, typical for
ions with less than half-filled $3d$-shell with the longest and
shortest Mn-O bond along the local $z$ and $x$ direction,
respectively \cite{Abragam70}.

The main result of this evaluation is the $E/D$-ratio, which we
improved in comparison to our previous estimate
\cite{Deisenhofer02} by the Q-band experiment. For $\vec{H}_{\rm
ext}$ applied along the $b$-axis the data are nearly
$T$-independent, whereas they clearly exhibit the CW behavior for
the other orientations. Regarding the equation for $g_b$, this is
only possible, if the factor $(3\zeta-1)$ is close to zero and
hence $\zeta$$\approx$$1/3$. This result is independent on the
value of the rotation angle $\gamma$. Only the absolute value of
$D$ directly depends on the choice of $\gamma$, which accounts for
the splitting of the resonance fields between $a$ and $c$
direction.

With the obtained $E/D$-ratio we now turn to the evaluation of the
linewidth data. A detailed derivation of the CF contributions to
the ESR linewidth in the cooperative JT distorted perovskite
structure accounting for the mutual rotations of the MnO$_6$
octahedra is presented in Ref.~\onlinecite{Kochelaev03}. These
theoretical considerations can be summarized in the following
formula
\begin{eqnarray}\label{dhfit}
\Delta H^{(\vartheta,\varphi)}(T) =& & \frac{T-T_{\rm CW}}{T}
[\Gamma_{\rm DM}(\infty) +   \\
&+& \left(1-\frac{T}{T_{\rm JT}}\right)^{2\beta}(\Gamma_{\rm
CF}(\infty)f_{\rm
reg}^{(\vartheta,\varphi)} \nonumber \\
&+& \Gamma_{\rm CFD}\left(\frac{T_{\rm N}}{6(T-T_{\rm
N})}\right)^{\alpha}f_{\rm div}^{(\vartheta,\varphi)})],\nonumber
\end{eqnarray}
where the first term describes the contribution $\Gamma_{\rm
DM}(\infty)$ of the DM interaction as introduced by Huber et
al.~\cite{Huber99} assuming that the exchange constants are
$T$-independent. This contribution is expected to survive the JT
transition and hence to determine the line broadening also at
$T>T_{\rm JT}$. The second and third term, $\Gamma_{\rm CF}$ and
$\Gamma_{\rm CFD}$, represent the regular and divergent CF
contributions, respectively. Only the latter diverges for
$T\rightarrow T_{\rm N}$ with an exponent $\alpha$, whereas both
terms decrease for $T\rightarrow T_{\rm JT}$ with a critical
exponent $2\beta$, with $\beta$ being the critical exponent of the
ZFS parameters $D$ and $E$. Note that both the assumption of $T$
independent exchange constants and the application of the
criticality at $T_{\rm JT}$ throughout the entire paramagnetic
$O^{\prime}$-phase can influence the value of the critical
exponent $\beta$ which therefore has to be regarded cautiously.
The angular factors $f_{\rm reg}^{(\vartheta,\varphi)}$ and
$f_{\rm div}^{(\vartheta,\varphi)}$ read
\begin{eqnarray}
f_{\rm reg}^{(\vartheta,\varphi)} &=& f_{\rm div}+(1+\zeta)^{2}
(1+\frac{3}{2}\sin^2\vartheta), \nonumber \\
f_{{\rm div}}^{(\vartheta,\varphi)} &=&\frac{1}{2}[1-3\zeta +
2\gamma(1+\zeta)]^2(1-\sin^2\vartheta\sin^2\varphi) \nonumber
\\&+& \frac{1}{2}[1-3\zeta -
2\gamma(1+\zeta)]^2(1-\sin^2\vartheta\cos^2\varphi),\nonumber
\end{eqnarray}
where $\vartheta$ and $\varphi$ are the polar and azimuthal angles
between $\vec{H}_{\rm ext}$ and the crystallographic $b$ and $c$
axes, respectively. An analogous calculation to the one presented
in \cite{Kochelaev03} showed that the DM interaction does not
exhibit any critical behavior at $T_{\rm N}$. To minimize the
number of fit parameters, we neglected here any angular dependence
of the DM contribution. In first approximation this is justified
by the observation that above $T_{\rm JT}$ the linewidth is
isotropic. However, generally an anisotropy of the DM contribution
can arise in the orbitally ordered state \cite{Deisenhofer02}. But
it consists itself at least in two contributions from different
Mn-O-Mn bond geometries, and its transformation to an isotropic
behavior on approaching $T_{\rm JT}$ needs further theoretical
considerations.
In the minimal model we omitted the divergent CF contribution
($\Gamma_{\rm CFD}$=0) and tried to find the best fit for the
linewidth data. The characteristic temperatures were kept fixed at
$T_{\rm CW}$=111~K and $T_{\rm JT}$=605~K and the rotation angle
in the $ac$ plane was set to $\gamma$=13$^{\circ}$. In addition
the parameter $\zeta$=0.37 was taken from the evaluation of the
$g$-values. So only three fit parameters remain, as there are the
regular part of the CF contribution $\Gamma_{\rm CF}(\infty)$, its
critical exponent $\beta$ at the JT transition, and the DM
contribution $\Gamma_{\rm DM}(\infty)$, which survives also at
temperatures $T>T_{\rm JT}$. A simultaneous fit (solid lines in
Fig.~\ref{dhxband}) of the $T$-dependence of $\Delta H$ is
satisfactorily performed above 200~K with $\Gamma_{\rm
DM}(\infty)$=1.0(1)~kOe, $\Gamma_{\rm CF}(\infty)$=0.57(2)~kOe,
and $\beta$=0.16(1).

Finally, we added the effect of the divergent CF contribution,
which allows to increase the difference between $\Delta H_a$ and
$\Delta H_c$ to lower temperatures, as observed in the experiment.
As shown in Fig.~\ref{dhxband} (dashed lines) a satisfactory
qualitative description of the data was obtained by $\Gamma_{\rm
CFD}$=10~kOe and $\alpha$=1.8 with fixed $T_N$=135~K. Comparison
with the Q-band data Fig.~\ref{qband} indicates that the magnetic
critical exponent below 200~K rather turns to the theoretically
expected lower value of $\alpha$=0.75 \cite{Kochelaev03}, which
however is not applicable at higher temperatures. The reason is
that the fit function is used for the whole temperature range, but
the derivation of the power law only holds for temperatures close
to $T_{\rm N}$. A further improvement may be achieved by taking
into account the symmetric anisotropic exchange interaction, which
is beyond the scope of this paper. In the following we concentrate
on the information obtained from the regular part of $\Delta H$.

The DM contribution $\Gamma_{\rm DM}(\infty)$=1.0~kOe determined
in the $O^{\prime}$-phase is lower than the one expected from
$\Delta H$$\approx$1.4~kOe in the $O$-phase. This discrepancy can
be explained by linewidth contributions of the CF due to the
dynamic JT effect present in the $O$-phase \cite{Sanchez03}.
Comparison with the regular CF contribution $\Gamma_{\rm
CF}(\infty)$=0.57~kOe allows to estimate the averaged value
$D_{\rm DM}$ of the DM interaction as defined in
Ref.~\onlinecite{Huber99}. The ratio of the linewidth
contributions equals the ratio of the respective second moments
approximated by $10(D_{\rm DM}/D)^2$$\approx$$\Gamma_{\rm
DM}(\infty)/\Gamma_{\rm CF}(\infty)$ \cite{Huber99}. With
$D$=0.6~K one obtains $D_{\rm DM}$$\approx$0.25~K. These values
are free from the uncertainty in the estimation of the exchange
frequency in the exchange-narrowed linewidth \cite{Kochelaev03},
because we used the $g$-values to determine the absolute values.
They are smaller than the values estimated earlier for
polycrystalline LaMnO$_3$ but their relative strength is in good
agreement with previous results \cite{Huber99,Tovar99}.

\begin{figure}[h]
\centering
\includegraphics[width=70mm,clip]{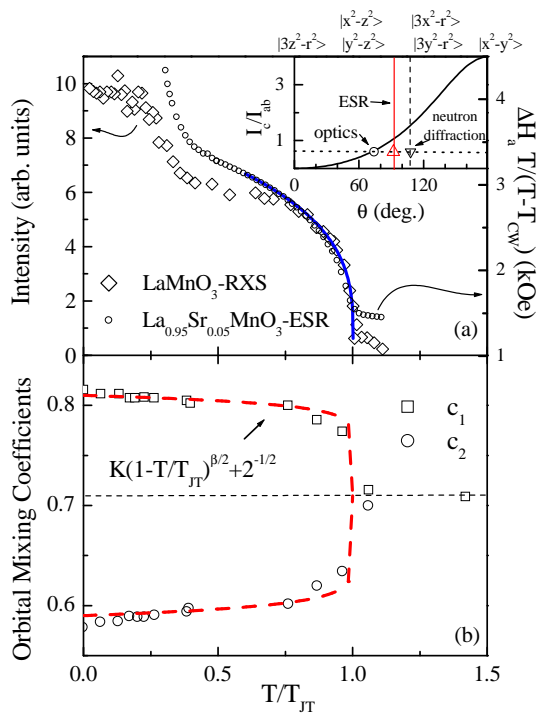}
\vspace{2mm} \caption[]{\label{orbxray} $T$-dependence of (a) the
reduced ESR linewidth in La$_{0.95}$Sr$_{0.05}$MnO$_{3}$ compared
to the RXS intensity of LaMnO$_{3}$ taken from \cite{Murakami98b}.
The solid line is a fit of the RXS intensity as described in the
text. Inset: $\Theta$-dependence of the normalized optical
spectral weight $I_{c}/I_{ab}$ following \cite{Tobe01} (solid
line) with the experimentally derived $\Theta$-values; (b)
$T$-dependence of $c_{1}$ and $ c_{2}$ determined by ND taken from
\cite{Rodriguez-Carvajal98}. The dashed lines were obtained using
$K(1-T/T_{\rm JT})^{\beta/2}$+ $2^{-1/2}$.}
\end{figure}

After having extracted the ZFS parameters we will now discuss the
consequences for OO in LaMnO$_3$:\\
In Fig.~\ref{orbxray}(a) the reduced ESR linewidth $\Delta
H_a$$\cdot$$T/(T-T_{\rm CW})$ for $\vec{H}_{\rm ext}$ parallel to
the $a$-axis, which bears the critical behavior on approaching
$T_{\rm JT}$ (see eq.~\ref{dhfit}), and the RXS intensity obtained
by Murakami \textit{et al.}\cite{Murakami98b} for LaMnO$_{3}$ are
shown to visualize the similarity of the two quantities on
approaching both $T_{\rm N}$ and $T_{\rm JT}$. Note that we used
the fixed value $T_{\rm CW}$=111~K of the $O^\prime$-phase leading
to somewhat lower values for the $O$-phase. It has been pointed
out that the RXS intensity close to $T_{\rm JT}$ is
$\propto$$(1-T/T_{\rm JT})^{2\beta}$ \cite{Ishihara02}, where
$\beta$ denotes the critical exponent of the OO parameter given by
the pseudo spin $\vec{T}$=$1/2(\sin \Theta,0,\cos \Theta)$
\cite{Maezono98,Ishihara02}. The angle
$\Theta$=2$\arctan(c_2/c_1)$ is a measure of the mixing of the
wave functions of the $e_g$-doublet in the ground state
$\psi_{g}$=$c_{1}|3z^{2}-r^{2}\rangle$+$c_{2}|x^{2}-y^{2}\rangle$.
To compare the results of RXS and ESR we fitted the RXS data in
the vicinity of the JT transition (650 K$\leq$$T$$\leq$$T_{\rm
JT}$=780~K) with such a critical behavior (solid line in
Fig.~\ref{orbxray}(a)) and obtained an exponent $\beta=0.16(1)$ in
agreement with the ESR linewidth. Considering that the ZFS
parameters can be denoted as $D$=$-3(\rho+\lambda
^{2}/\Delta)\cos\Theta$ and $E$=$-\sqrt{3}(\rho +\lambda
^{2}/\Delta )\sin \Theta$ with the spin-spin coupling $\rho$, the
SO coupling $\lambda$ and the $t_{2g}$-$e_g$ splitting energy
$\Delta$ \cite{Abragam70}, it is easy to identify  the relation
with the OO parameter $\vec{T}$.
Despite the lack of data in the critical regime,
Fig.~\ref{orbxray}(b) shows the orbital mixing coefficents $c_{1}$
and $c_{2}$ determined from a neutron diffraction (ND) study by
Rodriguez-Carjaval {\it et al.~}\cite{Rodriguez-Carvajal98} The
dashed lines were obtained by using $K(1-T/T_{\rm
JT})^{\beta/2}$+$2^{-1/2}$ ($T_{\rm JT}$=750~K) with a critical
exponent $\beta/2$ provided by $D$$\propto$$(c_{1}^{2}-c_{2}^{2})$
and $E$$\propto$$c_{1}\cdot c_{2}$. With $K_1$=0.10 and
$K_2$=-0.12 the data for $c_{1}$ and $ c_{2}$ can be well
described throughout the JT distorted phase. However, attempting
to describe the three data points in the critical regime rather
suggest $\beta$$\sim$0.3. Only a detailed structural study in the
critical regime will allow a direct confirmation of the critical
exponent for $c_{1}$ and $ c_{2}$, but $\beta/2$=0.08 seems to
yield a reasonable description.

The main result of this study is to estimate the angle
$\Theta$=$\sqrt{3}\arctan (E/D)$ and determine the type of OO by
using the obtained value $E/D$=0.37(1), which after taking into
account the transformation to a local coordinate system
\cite{Kochelaev03} results in $\Theta_{\rm ESR}$$\sim$$ 92^\circ$.
Another estimate has been obtained from ND via the orbital mixing
coefficents at room temperature $c_{1}$$\approx$ 0.8 and
$c_{2}$$\approx$0.6 resulting in $\Theta_{\rm ND}$=106$^\circ$
\cite{Rodriguez-Carvajal98}. This discrepancy cannot easily be
explained. However, Tobe \textit{et al.~}tried to explain the
anisotropy of the optical conductivity in LaMnO$_3$ on the basis
of a $p-d$ transition model by using the ratio
$I_{c}/I_{ab}=2(1-\cos\Theta)/(2+\cos\Theta)$ (see inset of
Fig.~\ref{orbxray}(a)) between the optical spectral weight with
the polarization within the ferromagnetic (FM) plane (there $ab$)
and along the AFM axis (there $c$).\cite{Tobe01} Their simple
model suggests a value $\Theta$$\sim$74$^\circ$ to describe the
experimental value of $I_{c}/I_{ab}$=0.6 at 10 K in the AFM
regime. Theoretically, only a slight decrease of $\Theta$ below
$T_{\rm N}$ is expected \cite{Okamoto,Sikora03} and therefore
above $T_{\rm N}$ $\Theta_{\rm ESR}$$\sim$92$^\circ$ yields a
better description of the anisotropic properties of LaMnO$_3$ in
the orbitally ordered state than $\Theta_{\rm
ND}$$\sim$106$^\circ$. Alejandro \textit{et al.}~derived a similar
$E/D$-value from ESR data in La$_{7/8}$Sr$_{1/8}$MnO$_{3}$
\cite{Alejandro01} which confirms our result. To the best of our
knowledge no $\Theta$-estimate has been obtained by RXS and it is
worthwhile to note that the RXS intensity becomes maximal for
$\Theta$=$\pi/2$ \cite{Ishihara02}, the value obtained by ESR.
Following \cite{Maezono98} $\Theta$=$\pi/2$ characterizes OO
stabilized by the SE processes in the FM bonds and the
electron-phonon coupling is small compared to the bandwidth.
Recent calculations suggest $\Theta$=$\pi/2$ at $T$=$T_{\rm N}$
and $\Theta$=83$^\circ$ at $T$=0 and confirm our findings
\cite{Sikora03}.

In summary, we were able to determine unambiguously the orbital
order of La$_{0.95}$Sr$_{0.05}$MnO$_{3}$. The evolution of the OO
parameter monitored by the RXS intensity shows an intriguing
similarity to the ESR linewidth. We found a mixing angle
$\Theta=92^\circ$ suggesting that OO is dominated by SE coupling
in agreement with theoretical predictions.

We thank M.V.~Eremin, T.~Kopp, and K.-H.~H{\"o}ck for fruitful
discussions. This work was supported by the BMBF under contract
No. 13N6917 (EKM) and partly by the DFG via the SFB 484 and
DFG/RFFI-project No.~436-RUS 113/566/0. B.I.K. and E.S. were
partially supported by CRDF via grant REC-007.


\end{document}